\documentclass[reprint,aps,prl]{revtex4-1}

\usepackage{amssymb}                            
\usepackage{fullpage}
\usepackage{graphicx}
\usepackage{caption}                            
\usepackage[squaren,Gray, cdot]{SIunits}        
\usepackage{amsmath}
\usepackage{natbib}
\usepackage{bm}
\usepackage{hyperref}
\usepackage{color}

\begin{document}

\title{A Sharp Phase Field Method}                       

\author{Alphonse Finel}
        \email{alphonse.finel@onera.fr}
\affiliation{Laboratoire d'Etude des Microstructures, ONERA, CNRS, Universit\'e Paris-Saclay, 92320 Ch\^atillon, France.}  
\author{Yasunori Yamada}
\affiliation{Institute for Materials Research, Tohoku University, Japan}  
\author{Yann Le Bouar}
\affiliation{Laboratoire d'Etude des Microstructures, ONERA, CNRS, Universit\'e Paris-Saclay, 92320 Ch\^atillon, France.}  
\author{Beno\^it Dabas}
\affiliation{Laboratoire d'Etude des Microstructures, ONERA, CNRS, Universit\'e Paris-Saclay, 92320 Ch\^atillon, France.}  
\author{Beno\^it Appolaire}
\affiliation{Laboratoire d'Etude des Microstructures, ONERA, CNRS, Universit\'e Paris-Saclay, 92320 Ch\^atillon, France.}  
\author{Tetsuo Mohri}
\affiliation{Institute for Materials Research, Tohoku University, Japan}

\date{\today}              
\begin{abstract}
Phase field modelling offers an extremely general framework to predict microstructural evolutions in complex systems. However, its computational implementation requires a discretisation scheme with a grid spacing small enough to preserve the continuous character of the theory. We present here a new formulation, which is intrinsically discrete, in which the interfaces are resolved with essentially one grid point with no pinning on the grid and an accurate rotational invariance, improving drastically the numerical capabilities of the method. We show that interfacial kinetic properties are reproduced with a high accuracy. Finally, we apply the model to a situation where conserved and non-conserved fields are coupled.

\end{abstract}



\maketitle

Phase Field Modelling (PFM) is intensively used in predicting microstructure evolutions in extremely diverse domains. The method consists in introducing a series of fields that represent the material properties of interest, such as atomic species concentrations and state of local order. These phase fields are used to identify locally the phase present at a given point but also the interfaces, which are represented by the rapid but smooth variations of the fields, the interface positions being implicitly given by the maxima of their gradients. The powerfulness of this concept is that it avoids the difficult problem of interface tracking and, most importantly, allows for any topological evolution of phase morphologies, such as interface instabilities, shape bifurcations, coagulation events, nucleation.  Based on these powerful capabilities, phase field methods enabled the simulation of complex evolution problems, such as solidification \cite{echebarria:2004}, solid-state transformations \cite{onuki:2001,chen:2002,finel:2010,Levitas:2010}, cracks propagation \cite{Henry:2004,Spatschek:2009}, dislocation dynamics \cite{rodney:2000,rodney:2003,shen:2003,ni:2008}, electromigration \cite{Bhate:2002,Park:2010}, fluid dynamics \cite{Nguyen:2010,Folch:1999} or biological processes \cite{Jamet:2008}.

Historically, the development of PFM may be traced back to van der Waals theory of diffuse interfaces \cite{vanderwaals:1893}, Landau theory of phase transitions \cite{Landau:1937a,Landau:1937b} and Cahn and Hilliard thermodynamic formulation of non uniform systems  \cite{cahn:1958}. As in these pioneering developments, the fundamental ingredient of PFM is an inhomogeneous  free energy density whose derivatives with respect to the phase fields provide driving forces for their dynamics. One of the reasons for the success of PFM is that, using simple symmetry arguments and the conserved or non-conserved characters of the fields, it is easy to develop free energy functionals and kinetic equations for complex situations where different phenomena are coupled.

However, a numerical implementation is required to integrate the kinetic equations, which are discretised on a computational grid. As the phase fields are assumed to vary continuously and in order to avoid artificial grid pinning, the grid spacing must be much smaller than the smallest internal length scale, i.e. the interfaces widths. This diffuse-interface constraint limits drastically the overall accessible linear dimensions or, conversely, increases dramatically the required computational time.

  The aim of this Letter is to introduce a Sharp Phase Field Method (S-PFM), in which interface widths may be as small as the grid spacing, without any pinning on the grid when the interfaces move, allowing to multiply the accessible linear dimensions by an order of magnitude or, conversely, to reduce the computational time by almost three orders of magnitude. 
  
\emph{Classical Phase Field Modelling} - For the sake of simplicity, we consider the simple case of a two-phase system in which the material properties may be represented by a single phase field $\phi(\vec r)$ which, away from any interface, may take only two different values, $\phi(\vec r)=0$ or $1$, depending on the phase present at point $\vec r$. The usual PFM formulation starts with a free energy functional $F = \int d^3r \{ g(\phi(\vec r)) + \frac12 \lambda \vert\vert \nabla \phi(\vec r)\vert\vert^2\}$ in which the field $\phi(\vec r)$ is continuously defined.  The free energy density $g(\phi)$ is a double-well potential, which in the present simple situation may be simply written as $g(\phi)=A\phi^2(1-\phi)^2$. The gradient term penalises spatial variations and, therefore, is responsible for the localised but diffuse character of the interfaces. As mentioned above, the numerical implementation of this continuous PFM formulation requires a grid spacing small enough to suppress grid pinning on the computational grid, i.e. to approximately recover the translational and rotational invariances formally lost by the discretisation scheme. Numerical experiences show that approximately 6 to 8 grid points across the interfaces are usually required.

\emph{1D Sharp Phase Field Model} -The starting point of our new formulation is a \emph{discrete} free energy functional. Considering first a one-dimensional situation, we introduce the following discrete free energy:
\begin{equation}
F =d \sum_n \{\,(g(\phi_n)) + \frac{\lambda}{2d^2}\,\vert\vert\tilde\nabla\phi_n\vert\vert^2\,\},
\label{discrete_free_energy}                       
\end{equation}
\noindent where $d$ is the grid spacing, $n$ labels the point of abscissa $x=nd$ and $\tilde\nabla\phi_n=\phi_n -\phi_{n-1}$ is a dimensionless discrete gradient. A discrete equilibrium profile between the two phases is given by $\frac{\partial F}{\partial \phi_n}=0$, i.e:
\begin{equation}
g'(\phi_n)-\lambda(\phi_{n+1}+\phi_{n-1}-2\phi_{n}) / d^2=0,
\label{differential_equation}                       
\end{equation}
\noindent with the boundary conditions $\lim_{n\rightarrow -\infty}\phi_n=0$ and $\lim_{n\rightarrow +\infty}\phi_n=1$. At this stage, the double-well potential $g(\phi)$ has not been defined. The key point is to identify a function $g(\phi)$ in such a way that, if $\phi_n=f(nd)$ is solution of Eq.~(\ref{differential_equation}), then $\phi_n=f(nd-x_0)$ is also a solution for \emph{any real} $x_0$. In that case, the interface energy will obviously be continuously invariant by translation, which is the key point to exactly suppress grid pinning. A necessary condition for the existence of a potential $g(\phi)$ that generates such interfaces is that Eq.~(\ref{differential_equation}) becomes an \emph{ordinary} differential equation which, even though its form will of course depend on the selected function $f(x)$, is invariant with respect to any real $x_0$. This will be achieved if $\phi_{n+1}$ and $\phi_{n-1}$ may be explicitly expressed in terms  of  $\phi_n$ and if $x_0$ does not appear in these expressions. It happens that the choice $f(x)=\frac{1+\tanh(\frac{x}{w})}{2}$ fulfils these requirements. Indeed, with
\begin{equation}
\phi_n=\frac{1+\tanh(\frac{nd-x0}{w})}{2},
\label{equilibrium_profile}                       
\end{equation}
we have the following property:
\begin{equation}
2\phi_{n\pm1}-1 = \frac{(2\phi_n-1)\,\pm\,\alpha}{1\,\pm\,(2\phi_n-1)\alpha},
\label{tanh_property}                       
\end{equation}
where the parameter $\alpha$ is defined by :
\begin{equation}
\alpha=\tanh(\frac{d}{w}).
\label{alpha}
\end{equation}
Using Eq.~(\ref{tanh_property}),  Eq.~(\ref{differential_equation}) becomes
\begin{equation}
g'(\phi) - \frac{\lambda}{d^2}\,\{\frac{1-\alpha^2}{1-\alpha^2(2\phi-1)^2}\,-\,1\}(2\phi-1)=0,
\label{differential_equation2}                       
\end{equation}
whose solution is
\begin{equation}
g(\phi) = \frac{\lambda}{4d^2}\{ \frac{\alpha^2-1}{\alpha^2}\log[1-\alpha^2(2\phi-1)^2] -(2\phi-1)^2\}.
\label{landau_potential}                       
\end{equation}
In brief, with this free energy density, the profile $\phi_n$ given in Eq.~(\ref{equilibrium_profile}) is an exact solution of the discrete equilibrium equation (\ref{differential_equation}) for \emph{any real} $x_0$. As a consequence, this discrete profile may be continuously translated along the $x$-axis without modifying the interface energy, which is a signature of absence of any pinning. The important point here is that there is no restriction on the parameter $w$ that enters into the phase field profile given in Eq.~(\ref{equilibrium_profile}). This parameter, which is directly linked to the interface thickness, may be chosen as small as we want, in particular smaller than the grid spacing $d$, in which case the interface width is even smaller than the grid spacing. 
\begin{figure}
\begin{minipage}[t]{1\linewidth}
\includegraphics[width=0.49\linewidth]{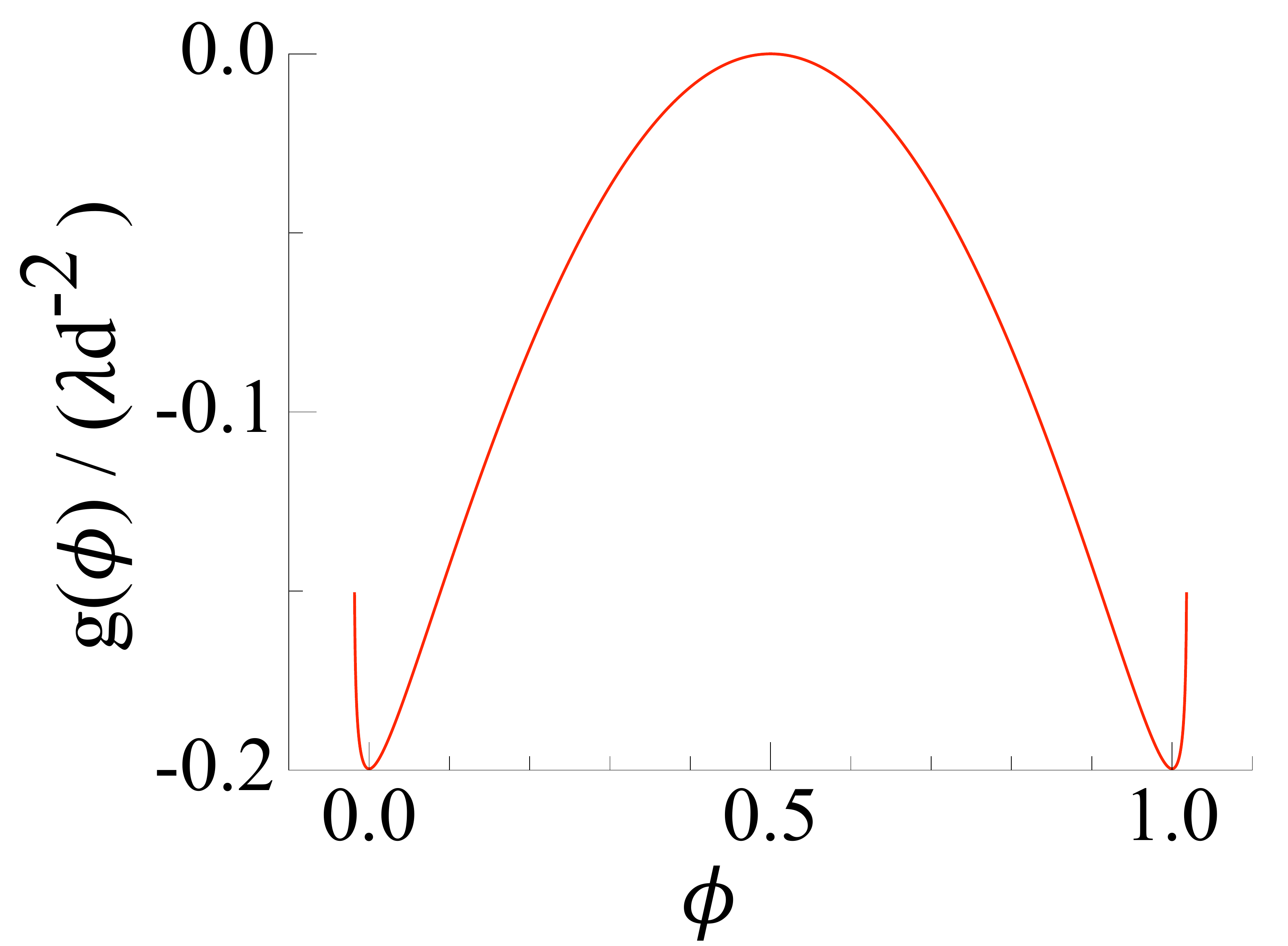}
\includegraphics[width=0.49\linewidth]{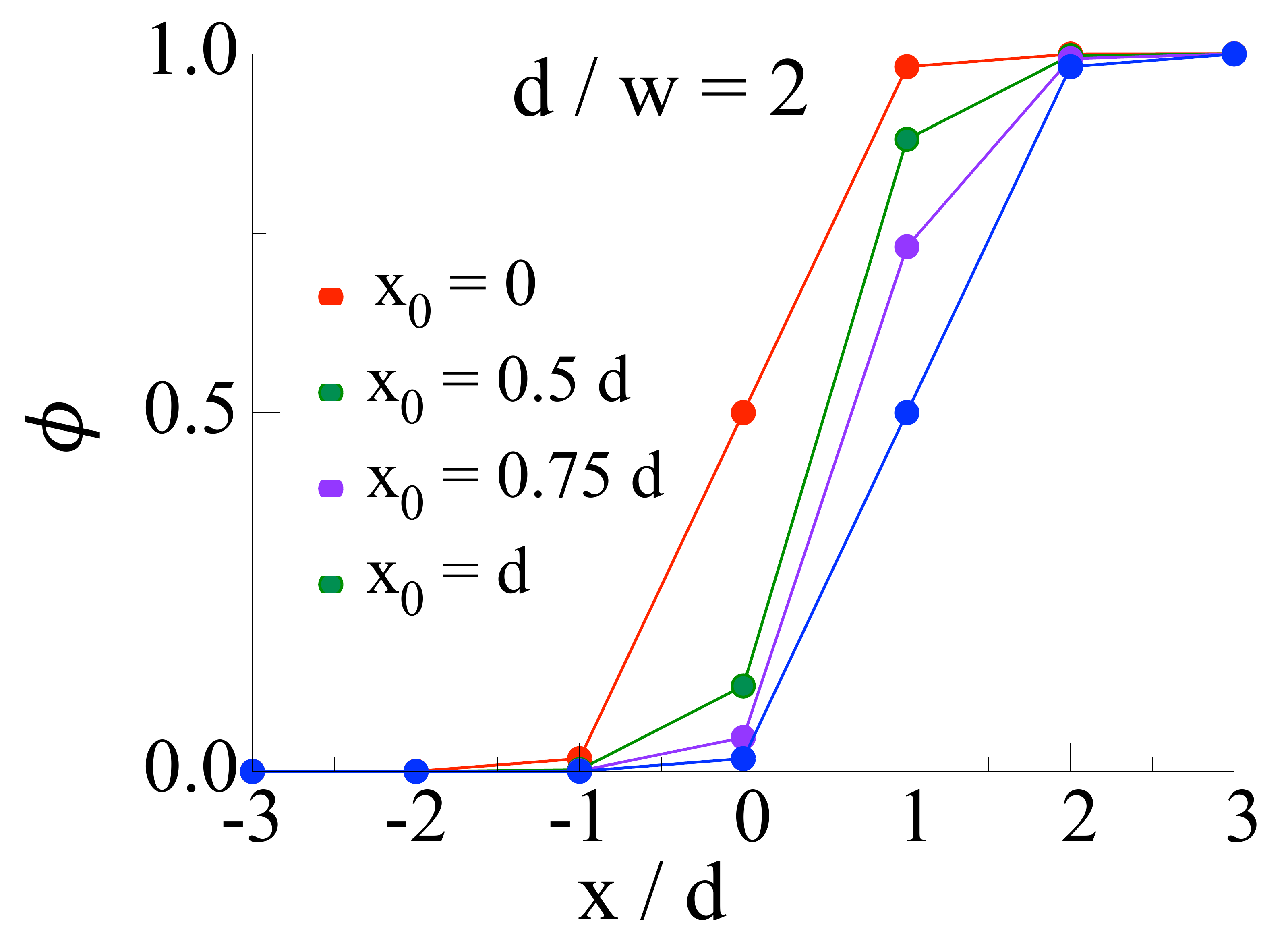}
\caption{Left : free energy density $g(\phi)$ as a function of $\phi$ for $d/w=2$ (see Eq.~(\ref{landau_potential})). Right: corresponding 1D phase field profiles for different shifting value $x_0$.}
 \label{fig1_potential_profiles}
\end{minipage}%
 \end{figure}
In order to illustrate the method, we consider the case $d/w=2$. The corresponding potential $g(\phi)$ is shown in Fig.~\ref{fig1_potential_profiles}. Besides its expected double-well form, we observe that the curvatures of $g(\phi)$ at the minima are much higher than at the top of the free energy barrier which separates the minima. This is in strong contrast with the situation observed when classical polynomial Landau potentials are used, such as the $\phi^2-\phi^4$ potential mentioned above. In fact, the sharpness of the potential wells increases exponentially with the selected ratio $d/w$ whereas the curvature at the top of the barrier stays finite. Equilibrium profiles associated to the free energy density given in Eq.~(\ref{landau_potential}) are shown in Fig.~\ref{fig1_potential_profiles} with different values of the shifting quantity $x_0$ (see Eq.~(\ref{equilibrium_profile})). As expected, we observe that these profiles display sharp interfaces that are resolved with essentially only one grid point. Most importantly, we stress that, even though the discrete values reached by the  phase field through the interfaces vary drastically with $x_0$, we checked that, up to fourteen digits, all the profiles generate exactly the same interface energy $\sigma\simeq0.44263\,\lambda d^{-1}$, which confirms the absence of any  pinning on the grid.


\emph{3D Sharp Phase Field Model} - The previous analysis results in an exact formulation for a PFM modelling where the interfaces width may be as small as we want, even smaller that the grid spacing and, yet, with no pinning on the grid. Now, in order to tackle problems of interest, we need to extend this sharp interface approach to 3-dimensional situations, keeping its discrete character. However, it is straightforward to show that it is impossible to construct a free energy density $g(\phi)$ in such a way that it generates flat interfaces with no pinning along more than one direction (except for obvious degeneracies due to the grid symmetries). In other words, it is impossible to fully recover the translational and rotational invariances lost by the introduction of a discrete 2D or 3D computational grid. Therefore, we proceed as follows. We first select a lattice plane family on the computational grid, referred to by its Miller indices $(h_1k_1l_1)$ in a reference basis, and construct a free energy potential $g(\phi)$ that generates an interface energy $\sigma(h_1k_1l_1)$ translationally  invariant along the directions perpendicular to these planes. Next, in order to approximately correct for the otherwise broken rotational invariance, we select two other lattice plane families, noted  $(h_2k_2l_2)$ and $(h_3k_3l_3)$, and proceed in such a way that the interface energies of the corresponding interfaces match exactly $\sigma(h_1k_1l_1)$. More precisely, as they cannot be invariant by translation, the interface energies $\sigma(h_2k_2l_2)$ and $\sigma(h_3k_3l_3)$ are defined as the average of the extremum values the interface energies reach when the corresponding interfaces move along their perpendicular directions. Applicability of the condition of equality of three interface energies requires the identification of two degrees of freedom. For that purpose, we extend the gradient term that appears in the discrete free energy to the 3rd neighbour shells and write :
\begin{equation}
\frac{F}{V_0}=\sum_{\vec r}\{g(\phi(\vec r) )+ \frac{\lambda}{2}\sum_{i=1}^3\gamma_i\frac{\nu_i}{d_i^2}\sum_{k=1}^{m_i}\vert\vert\phi(\vec r + \vec r_i(k))-\phi(\vec r)\vert\vert^2 \},
\label{3D_free_energy}                       
\end{equation}
where $V_0$ is the volume per node of the discrete lattice,  $\vec r$ runs over the nodes of the lattice, index $i$ labels the $i^{th}$ neighbour shell and, for a given shell $i$, index $k$ runs over the $m_i$ neigbouring nodes $\{\vec r_i(1)...\vec r_i(m_i)\}$ that constitute this shell. The coefficient $\nu_i=3/m_i$ corrects for the multiplicity of shell $i$ and $d_i$ is the length of a $i^{th}$ neighbour pair. The parameter $\gamma_i$ represents the weight of the $i^{th}$ neighbour shell in the gradient term, with the constraint that $\sum_{i=1}^3 \gamma_i=1$. In the continuum limit, the sum of the three gradient terms of Eq.~(\ref{3D_free_energy}) converges to $\vert\vert \nabla\phi\vert\vert^2$. The discrete free energy formulation given in Eq.~(\ref{3D_free_energy}) is  general and may be used for any computational grid. Its application to a specific grid requires only, for each shell $i$ used in the gradient term, the identification of its neighbouring nodes $\{\vec r_i(1)...\vec r_i(m_i)\}$. A cubic grid is often used. Here, we use  a face-centered cubic (FCC) grid \footnote{Our model may be extended to mechanical fields (strain and stress) that relax instantaneously. In the present sharp interface context, a very stable elastic solver is then required, which is easily done on an FCC grid.}.
As explained above, we want the free energy density $g(\phi)$ that appears in Eq.~(\ref{3D_free_energy}) be such that the interface energy $\sigma(h_1k_1l_1)$ associated to planes of type $(h_1k_1l_1)$ is strictly translationally invariant. Following the procedure used above, this leads to the expression
\begin{align}
g(\phi)&= \frac{\lambda}{4}\,\sum_{i=1}^3\,\gamma_i\,\frac{\nu_i}{d_i^2}\sum_{k=1}^{m_i}\{ \frac{\alpha_i(\vec r_i(k))^2-1}{\alpha_i(\vec r_i(k))^2} \nonumber \\
&\log[1-\alpha_i(\vec r_i(k))^2(2\phi-1)^2]-(2\phi-1)^2  \},
\label{3D_free_energy_density}                       
\end{align}
with the coefficients $\alpha_i(\vec r_i(k))$ given by 
\begin{equation}
\alpha_i(\vec r_i(k))=\tanh(\frac{\vec r_i(k) . \vec u }{w}),
\label{coef_alpha}                       
\end{equation}
where $\vec u$ is a unit vector perpendicular to the planes $(h_1k_1l_1)$. In Eq.~(\ref{3D_free_energy_density}), the second summation is restricted to $k$-values for which $\alpha_i(\vec r_i(k))$ is non zero. As for the 1D case discussed above, the parameter $w$, which controls the width of the interfaces $(h_1k_1l_1)$, may be chosen as small as we want.

Now, we proceed with the optimisation scheme proposed above. Once a translationally invariant family $(h_1k_1l_1)$ and its companion families $(h_2k_2l_2)$ and $(h_3k_3l_3)$ have been chosen, we must identify ponderation coefficients $\gamma_i$ such that the interface energies verify $\sigma(h_1k_1l_1)=\sigma(h_2k_2l_2)=\sigma(h_3k_3l_3)$. Of course, the outcome of this procedure will depend on which lattice family is first selected for receiving the translational invariance. Two different criteria may be used. First, we may argue that a good way to simultaneously optimise for the translational and rotational invariances is to select a plane family that displays the highest possible degeneracy with respect to the symmetries of the computational grid, which, for a grid with cubic symmetry, is equal to 24. The second criterion concerns the inter-reticular distance. Indeed, pinning effects on the internal energy of flat interfaces should increase with the inter-reticular distance. Therefore, the remaining pinning effects on the interfaces that have not been selected for the translational invariance will be minimised if the plane family that receives this invariance displays a large inter-reticular distance. The first criteria is fulfilled by any family whose Miller indices are all different and, among those, the best candidate is family $(135)$ (Miller indices are expressed in the orthogonal basis defined by the edges of the unit FCC cube) because it also maximises the inter-reticular distance. On the other hand, the second criteria is fulfilled by family $(111)$, which corresponds to the planes with the highest two-dimensional packing.

We now present the numerical results of the optimisation procedure for a ratio between the size $d$ of the unit FCC cube and the characteristic interface length scale $w$ fixed to $d/w=3$. The choices $(135)$ and $(111)$ for the translationally invariant families $(h_1k_1l_1)$ have been analysed. For each of these choices, two other plane families have been selected and the ponderation coefficients $\gamma_i$, of which only two are independent, have been optimised in the sense defined above. The results are displayed in Tab.~\ref{table1}.
\begin{table}[b]
\begin{center}
\begin{tabular}{|c|c|c|c|c|c|}
\hline
$(h_1k_1l_1)$ & $(h_2k_2l_2)$ & $(h_3k_3l_3)$ & $\gamma_2$ & $\gamma_3$ & $\tilde \sigma = \sigma /(\lambda d^{-1})$ \\
\hline
(135) & (111) & (200) & 0.1785 & 0.2935 &0.6671 \\
(111) & (200) & (220) & 0.1736 & 0.2545 &0.6720 \\
\hline
\end{tabular}
\caption{Optimised coefficients $\gamma_2$ and $\gamma_3$ for the 2nd and 3rd neighbour contributions ($d/w=3$)}
\label{table1}
\end{center}
\end{table}
The quality of the sharp interface modelling associated to the parametrisations shown in Tab.~\ref{table1} is now tested through the simulation of the growth of a single precipitate.  For that, we simply use a non-conserved dissipative dynamics on the phase field,
\begin{equation}
\frac{\partial \phi}{\partial t}=-L\frac{\delta F}{\delta \phi},
\label{coef_L}                       
\end{equation}
\begin{figure}
\begin{minipage}[t]{1\linewidth}
\includegraphics[width=0.49\linewidth]{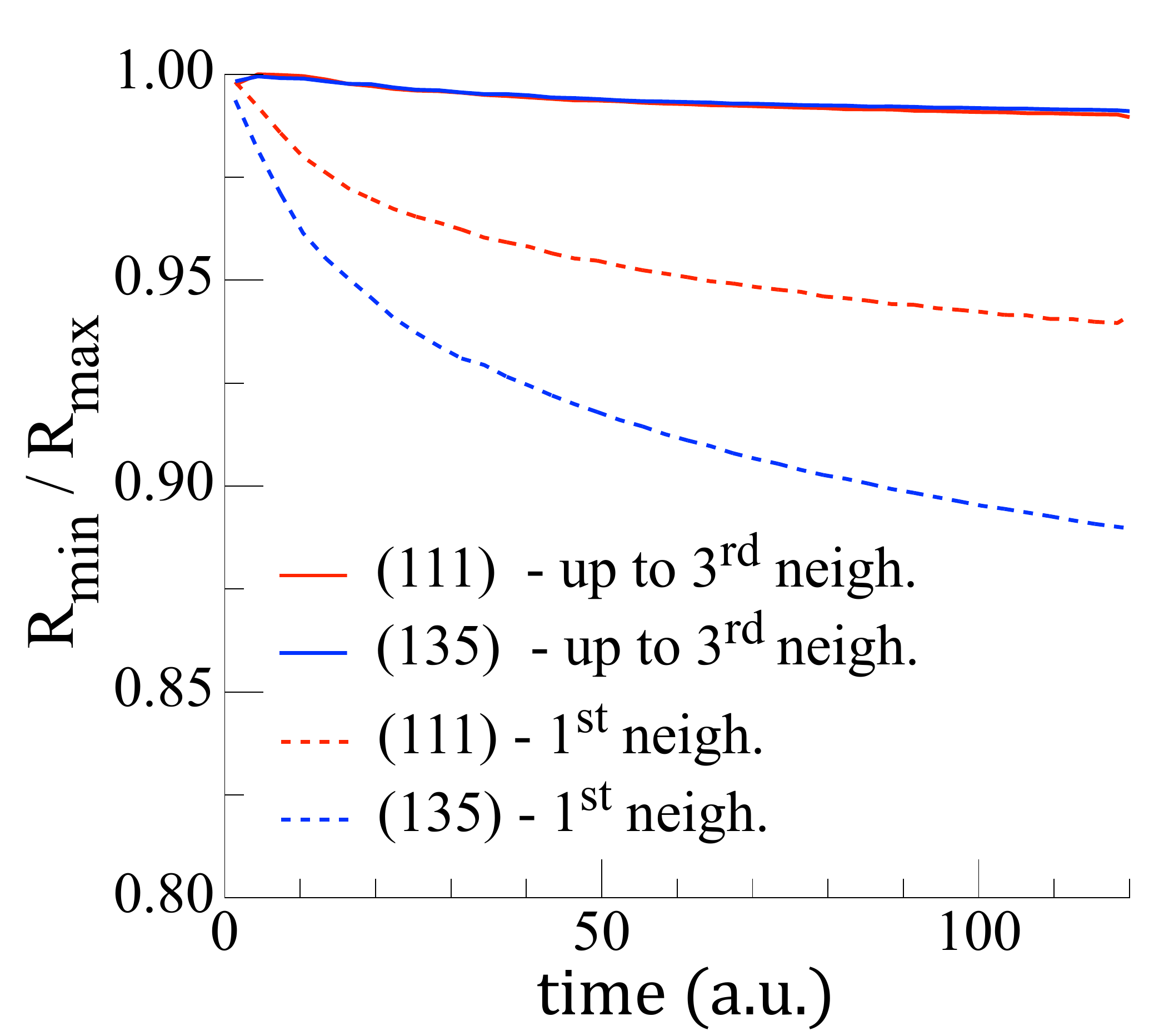}
\includegraphics[width=0.49\linewidth]{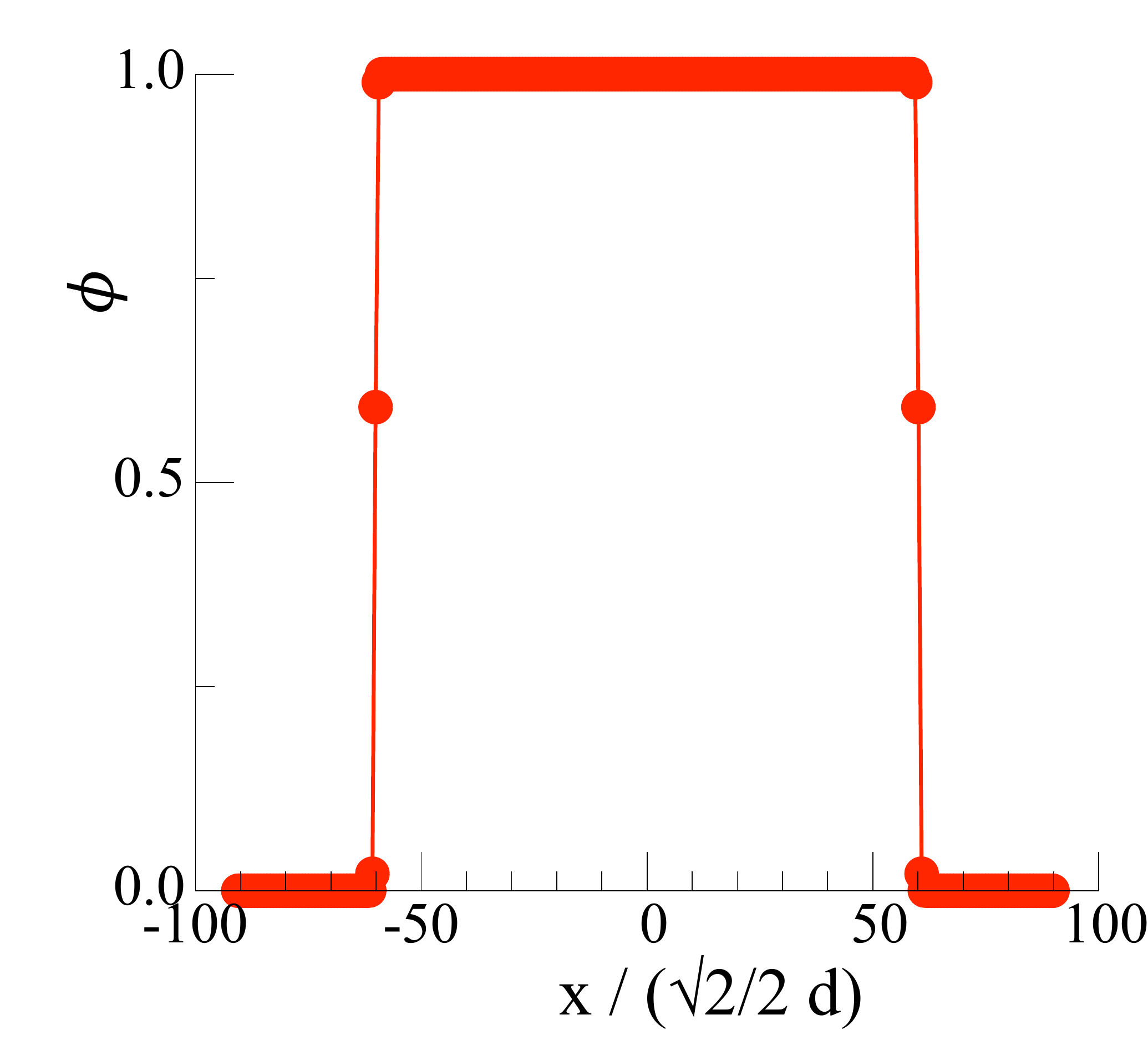}
\caption{Left: sphericity as function of time for a growing precipitate in a $256^3$ simulation box with gradient terms limited to 1st neighbours (dotted lines) or extended to 3rd neighbours (full lines). Right: phase field profile along a middle [110] line, obtained with the optimisation scheme (111).}
\label{fig3_sphericity}
\end{minipage}
\end{figure}
\begin{figure}
\begin{minipage}[t]{1\linewidth}
\includegraphics[width=0.59\linewidth]{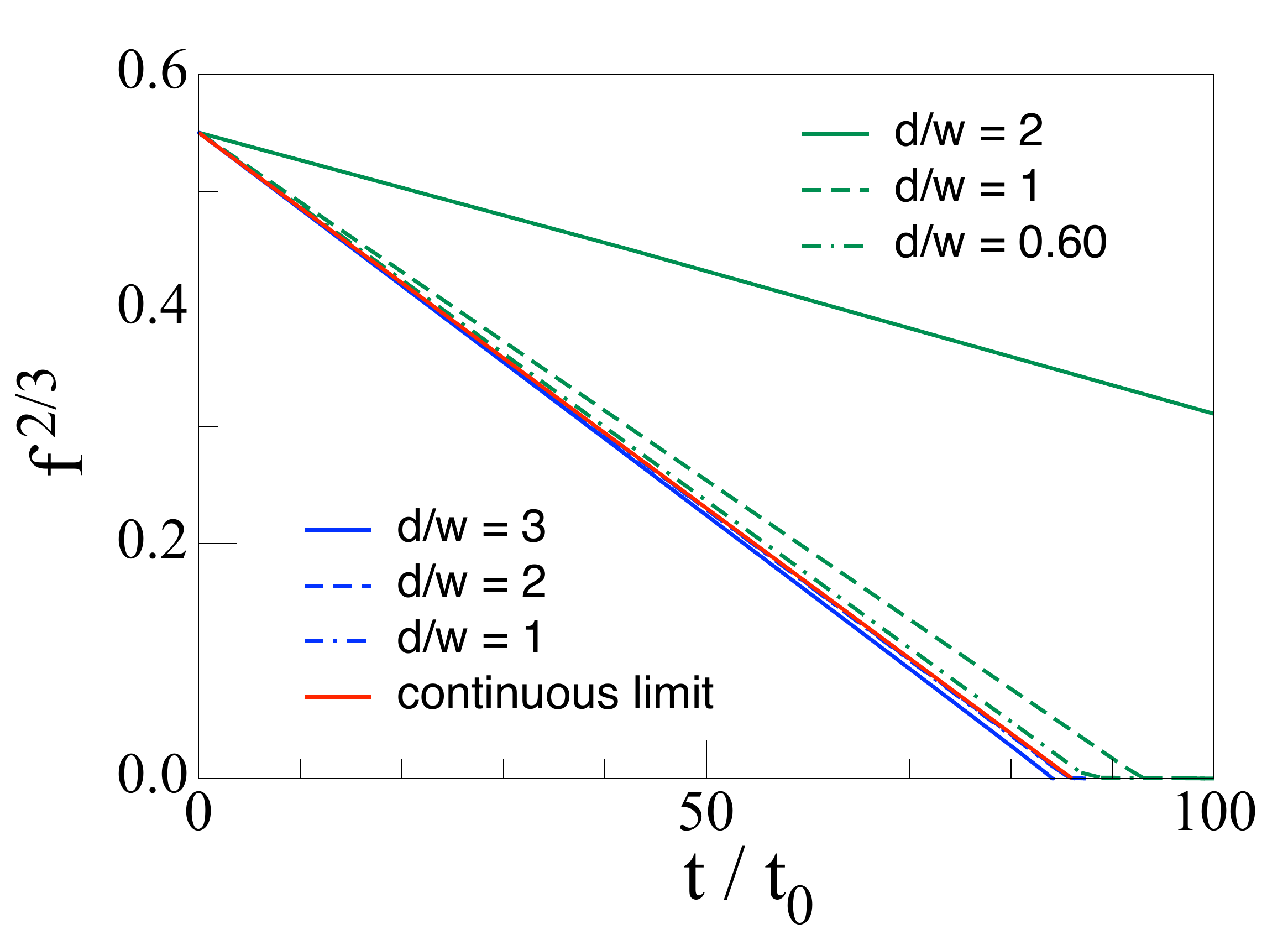}
\includegraphics[width=0.39\linewidth]{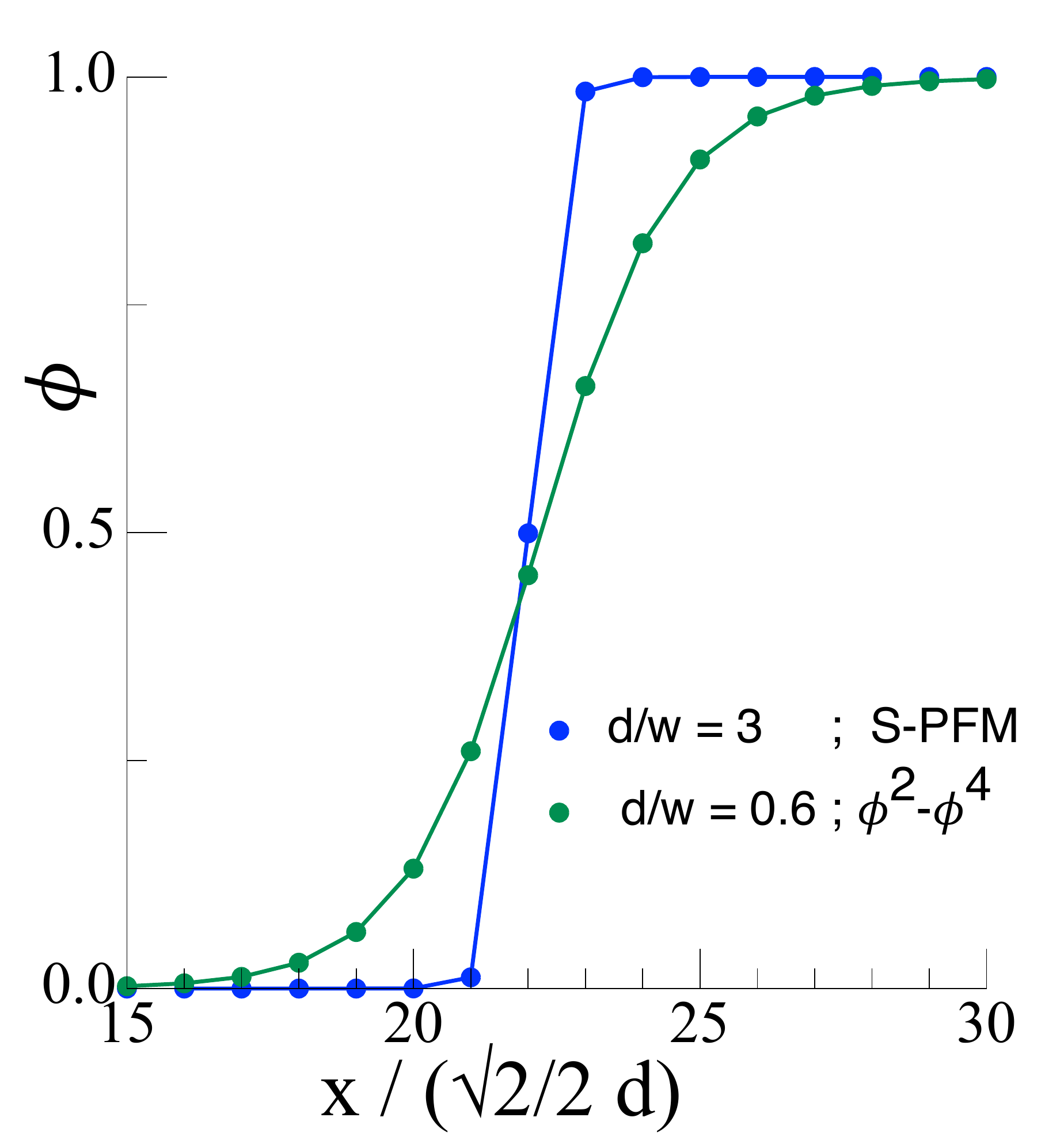}
\caption{Comparison between S-PFM (blue) and classical PFM (green). Left: volume fraction of a shrinking precipitate for different values of the ratio $d/w$ (time unit is $t_0=d^2/\mu \sigma)$. Right: profiles for a precision on the velocity of $1.7\%$.}
\label{kinetic_properties}
\end{minipage}%
 \end{figure}
where the total free energy $F$ of Eq.~(\ref{3D_free_energy}) is supplemented by a term of the form $-\Delta f h(\phi)$, where $h(\phi)=3\phi^2-2\phi^3$ is an interpolation function that favours the growth of domains where the field $\phi$ reaches 1. The results for an initially small spherical domain are presented in Fig.~\ref{fig3_sphericity}, where we show the time evolution of the sphericity indicator $R_{min}/R_{max}$, where $R_{min}$ ($R_{max}$) is the smallest (largest) distance between any grid point outside (inside) the precipitate and its center. In order to appreciate the benefit of using the optimisation procedure, we also display the results obtained with gradient terms limited to the 1st neighbour shell, in which case the sharp interface modelling is simply limited to the selection of the translationally invariant plane family $(h_1k_1l_1)$. The time and length scales analysed here are large as, in its final state, the precipitate periphery almost reaches the limits of the simulation box. We observe in Fig.~\ref{fig3_sphericity} that, with gradient terms limited to 1st neighbours, the precipitate sphericity is rapidly deteriorated, even though the results obtained with $(h_1k_1l_1)=(111)$, which maximises the inter-reticular distance, are clearly better than those obtained with $(h_1k_1l_1)=(135)$. When the optimisation procedure with gradient terms extended up to the 3rd neighbour shell is used, the results are much better: even though it increases with time, the sphericity loss remains smaller than $1\%$ for the two $(h_1k_1l_1)$ choices which, therefore, lead to simulation data that are indistinguishable. We note that this almost perfect sphericity is achieved even though the precipitate interface stays extremely sharp, as shown in Fig.~\ref{fig3_sphericity}, where it is seen that the interface is resolved with only one grid point. We mention that simulations with $d/w=2$ leads to an even better sphericity, with a loss of the order of $0.1\%$ and interfaces still resolved with essentially one point.

Next, we analyse the ability of the model to reproduce kinetic properties. For that, the links between the parameters of the theory, i.e. prefactor $\lambda$ (Eq.~(\ref{3D_free_energy})) and mobility $L$ (Eq.~(\ref{coef_L})), and materials properties are needed. For any grid spacing $d$ and parameter $w$, these are easily shown to be $\lambda = \sigma d / \tilde \sigma$ and $L=\mu/3\omega$, where $\sigma$ is the interface energy, $\mu$ the kinetic interfacial coefficient and $\tilde\sigma$ the numerically computed dimensionless interface energy (see Tab.~\ref{table1}). The results are shown in Fig.~\ref{kinetic_properties}, where we display the volume fraction as a function of time of a curvature driven shrinking precipitate. In contrast to classical PFM, S-PFM closely reproduces the exact continuum limit even for large $d/w$ (the results for $d/w=1$ and $d/w=2$ are indistinguishable from the limit). The improvement is clearly seen in Fig.~\ref{kinetic_properties}, where we display the phase field profiles of the S-PFM and classical  implementations that both reproduce the interface kinetics within a precision of $1.7\%$ : S-PFM requires only one grid point in the interface, whereas at least 8 points are needed for the classical formulation.
 \begin{figure}
\begin{minipage}[t]{1\linewidth}
\includegraphics[width=0.49\linewidth]{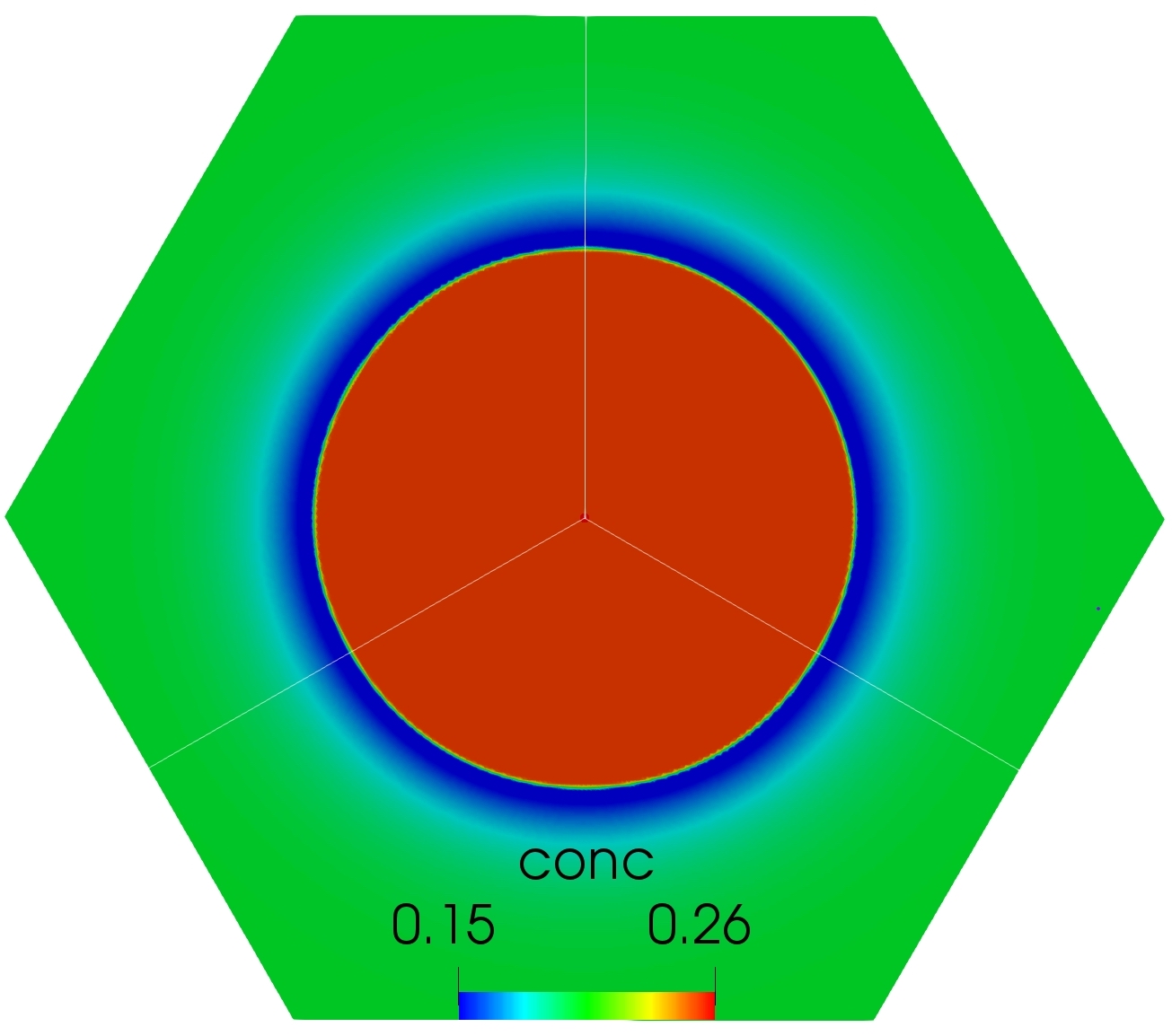}
\includegraphics[width=0.49\linewidth]{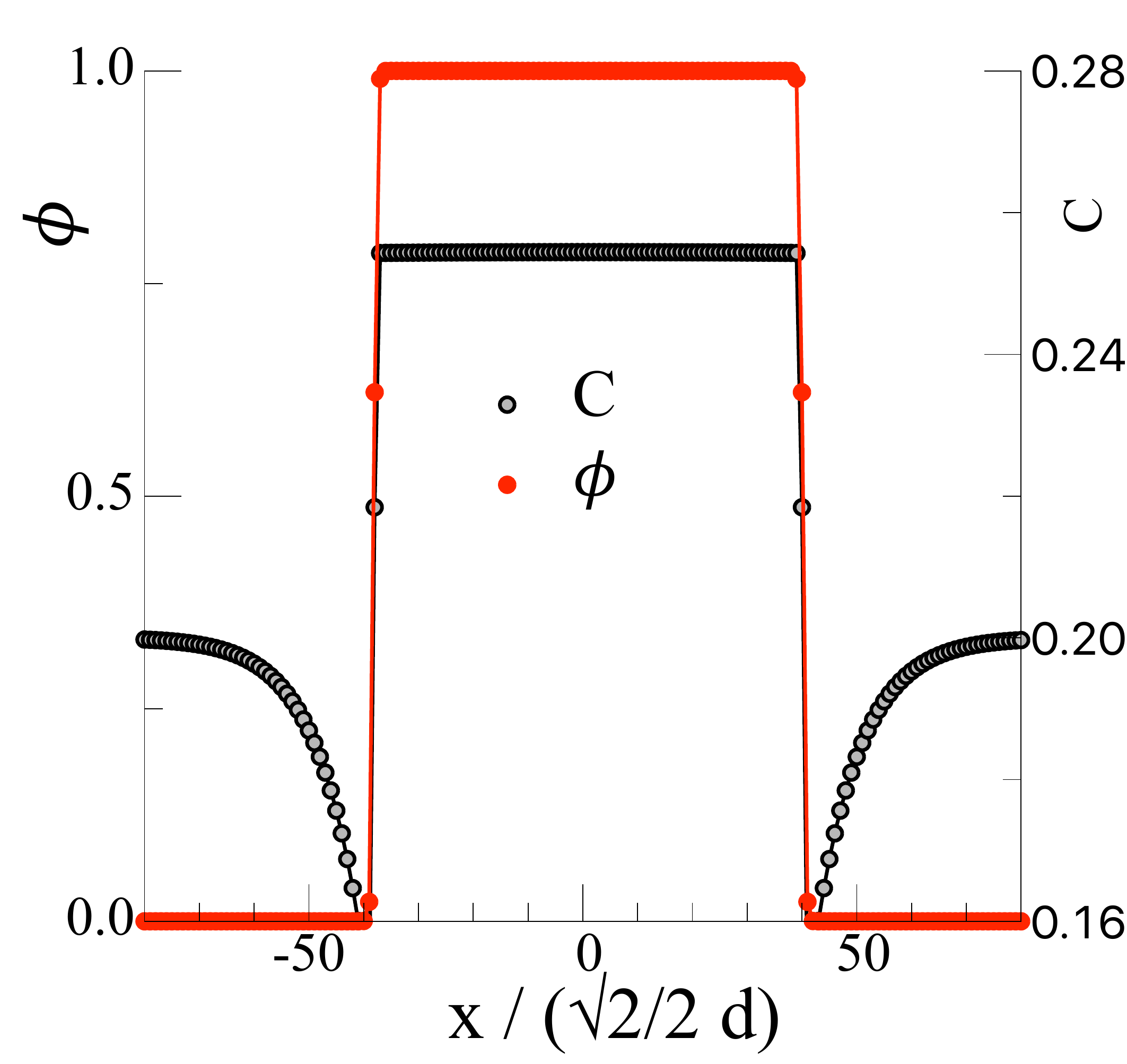}
\caption{Growing precipitate in a supersatured matrix. Left, (111) cut through the center of the $256^3$ simulation box; right, concentration and phase field profiles along a middle [110] line.}
\label{coupling_conserved_field}
\end{minipage}
 \end{figure}

In most studies, modelling of materials of interest requires more than one field. A typical example is a binary system in which precipitates differ from the matrix by local atomic order \emph{and} atomic concentrations. In that case, a supplemental field, here the concentration of one of the constituent, is required. We therefore extended our sharp interface model to such situations. Specifically, we supplemented the free energy of Eq.~(\ref{3D_free_energy}) with a term that makes the concentration field $c$ reach an equilibrium value for $\phi=0$ that differs from the one reached for $\phi=1$. Most importantly, in order to let the interface properties be controlled by the field $\phi$, the only gradient terms of the theory must stay the ones already introduced. This is a necessary condition to keep the sharp interface character of our model. The results are shown in Fig.~\ref{coupling_conserved_field}, where we display a growing precipitate in a supersaturated matrix. We note that the interface is still very sharp, even though we observed the expected smooth concentration dip in front of the growing precipitate. 

In brief, we have presented a Sharp Phase Field Model in which interfaces are numerically resolved with essentially one grid point, with no pinning on the grid and an accurate rotational invariance. We have shown that the model reproduces accurate kinetic interfacial properties. Finally, we have shown that the model can be used in situations that involve simultaneously non-conserved and conserved fields.


\small
\bibliographystyle{apsrev4-1.bst}

\end{document}